\documentclass[lettersize,journal]{IEEEtran}
\usepackage{amsmath,amsfonts}
\usepackage{algorithmic}
\usepackage{algorithm}
\usepackage{array}
\usepackage[caption=false,font=normalsize,labelfont=sf,textfont=sf]{subfig}
\usepackage{textcomp}
\usepackage{stfloats}
\usepackage{url}
\usepackage{verbatim}
\usepackage{graphicx}
\usepackage{cite}
\usepackage[english]{babel}
\usepackage{caption}
\usepackage{makecell}
\usepackage{todonotes}
\usepackage{soul}
\usepackage{pdflscape}

\usepackage{amsmath}
\usepackage{amsfonts}
\usepackage{amssymb}
\usepackage{graphicx}
\usepackage{xspace}
\usepackage{xcolor}
\usepackage{hyperref}
\usepackage{colortbl}
\usepackage{dsfont}

\usepackage{todonotes}
\usepackage{etoolbox}
\usepackage{xargs}
\usepackage{multirow}

\hypersetup{
    colorlinks = true,
    linkcolor=red,
    citecolor=red,
    urlcolor=blue,
    linktocpage 
}

\newcommand{\fmriprep}{\emph{fMRIPrep}\xspace}
\newcommand{\fwhm}{\textsc{FWHM}}

\newtoggle{RRandRS}
\togglefalse{RRandRS} 

\newlength{\figureWidthUM}
\newlength{\vspaceIndex}
\newlength{\rowSpace}

\newcommandx{\uncertaintyMap}[5][5=false]{%
  \setlength{\figureWidthUM}{0.15\paperheight}
  \ifstrequal{#5}{true}{\setlength{\vspaceIndex}{-45pt}}{\setlength{\vspaceIndex}{25pt}}
  \begin{tabular}[b]{@{}c@{}}
    #2\vspace*{\vspaceIndex}
  \end{tabular}
  \begin{tabular}[t]{@{}c@{}}
    \ifstrequal{#5}{true}{IEEE (T1 intensity)}{} \\
    \includegraphics[width=\figureWidthUM]{figures/sig/#1mm/ieee_#3_#4.pdf}
  \end{tabular}
  \begin{tabular}[t]{@{}c@{}}
    \ifstrequal{#5}{true}{RR (significant bits)}{} \\
    \includegraphics[width=\figureWidthUM]{figures/sig/#1mm/rr_#3_#4_sig.pdf}
  \end{tabular}
  \begin{tabular}[t]{@{}c@{}}
    \ifstrequal{#5}{true}{RS (significant bits)}{} \\
    \includegraphics[width=\figureWidthUM]{figures/sig/#1mm/rs_#3_#4_sig.pdf}
  \end{tabular}
  \iftoggle{RRandRS}{%
    \begin{tabular}[t]{@{}c@{}}
      \ifstrequal{#5}{true}{RR+RS (significant bits)}{} \\
      \includegraphics[width=\figureWidthUM]{figures/sig/#1mm/rr.rs_#3_#4_sig.pdf}
    \end{tabular}
  }{}
}

\newcolumntype{P}[1]{>{\centering\arraybackslash}p{#1}}
\newcolumntype{M}[1]{>{\centering\arraybackslash}m{#1}}

\newcommandx{\uncertaintyMapDiscrete}[5][5=false]{%
  \setlength{\figureWidthUM}{0.16\paperheight}
  \ifstrequal{#5}{true}{\setlength{\vspaceIndex}{-45pt}}{\setlength{\vspaceIndex}{15pt}}
  \ifstrequal{#5}{true}{
    \begin{tabular}[t]{@{}P{\figureWidthUM}@{}P{\figureWidthUM}@{}P{\figureWidthUM}@{}}
      IEEE (T1 intensity) & RR (significant bits) & RS (significant bits) 
    \end{tabular}
  }{}
  \begin{tabular}[t]{m{3pt}P{\figureWidthUM}@{}P{\figureWidthUM}@{}P{\figureWidthUM}@{}}
    {#2 \vspace*{0.25\figureWidthUM}} & \includegraphics[width=\figureWidthUM]{figures/sig/#1mm/ieee_#3_#4.pdf} &
    \includegraphics[width=\figureWidthUM]{figures/sig/discrete/#1mm/rr_#3_#4_sig_discrete.pdf} &
    \includegraphics[width=\figureWidthUM]{figures/sig/discrete/#1mm/rs_#3_#4_sig_discrete.pdf}
  \end{tabular}
  \iftoggle{RRandRS}{%
    \begin{tabular}[t]{@{}c@{}}
      \ifstrequal{#5}{true}{RR+RS (significant bits)}{} \\
      \includegraphics[width=\figureWidthUM]{figures/sig/discrete/#1mm/rr.rs_#3_#4_sig_discrete.pdf}
    \end{tabular}
  }{}
}
    
\newcommandx{\uncertaintyMapDiscreteAll}[1]{%
  \centering
  \setlength{\rowSpace}{-24pt}
  \uncertaintyMapDiscrete{#1}{1}{ds000256}{sub-CTS201}[true] \\
  \vspace*{\rowSpace}
  \uncertaintyMapDiscrete{#1}{2}{ds000256}{sub-CTS210} \\
  \vspace*{\rowSpace}
  \uncertaintyMapDiscrete{#1}{3}{ds001600}{sub-1} \\
  \vspace*{\rowSpace}
  \uncertaintyMapDiscrete{#1}{4}{ds001748}{sub-adult15} \\
  \vspace*{\rowSpace}
  \uncertaintyMapDiscrete{#1}{5}{ds001748}{sub-adult16} \\
  \vspace*{\rowSpace}
  \uncertaintyMapDiscrete{#1}{6}{ds001771}{sub-36} \\
  \vspace*{\rowSpace}
  \uncertaintyMapDiscrete{#1}{7}{ds002338}{sub-xp201} \\
  \vspace*{\rowSpace}
  \uncertaintyMapDiscrete{#1}{8}{ds002338}{sub-xp207} \\
}

\title{A numerical variability approach to results stability tests and its application to neuroimaging}
\author{\IEEEauthorblockN{Yohan Chatelain\IEEEauthorrefmark{1}, Lo\"ic Tetrel\IEEEauthorrefmark{2}, Christopher J. Markiewicz\IEEEauthorrefmark{3}, Mathias Goncalves\IEEEauthorrefmark{3}, Gregory Kiar\IEEEauthorrefmark{6},\\ Oscar Esteban\IEEEauthorrefmark{3}\IEEEauthorrefmark{5},  Pierre Bellec\IEEEauthorrefmark{2}\IEEEauthorrefmark{4}, Tristan Glatard\IEEEauthorrefmark{1}\vspace*{0.2cm}}

\IEEEauthorblockA{\IEEEauthorrefmark{1}Department of Computer Science and Software Engineering\\ Concordia University, Montreal, Quebec, Canada.}

\IEEEauthorblockA{\IEEEauthorrefmark{2} Centre de recherche de l'Institut Universitaire de Gériatrie\\ de Montréal (CRIUGM), Montréal, Québec, Canada.}

\IEEEauthorblockA{\IEEEauthorrefmark{3} Department of Psychology, Stanford University, Stanford, CA, USA.}

\IEEEauthorblockA{\IEEEauthorrefmark{4} Department of Psychology, Université de Montréal, Montréal, Québec, Canada.}

\IEEEauthorblockA{\IEEEauthorrefmark{5} Department of Radiology, Lausanne University Hospital\\ and University of Lausanne, Switzerland.}

\IEEEauthorblockA{\IEEEauthorrefmark{6} Child Mind Institute, New York City, NY, USA.}

}

\begin{document}
\maketitle

\begin{abstract}
  Ensuring the long-term reproducibility of data analyses requires results stability tests to verify that analysis results remain within acceptable variation bounds despite inevitable software updates and hardware evolutions. This paper introduces a numerical variability approach for results stability tests, which determines acceptable variation bounds using random rounding of floating-point calculations. By applying the resulting stability test to \fmriprep, a widely-used neuroimaging tool, we show that the test is sensitive enough to detect subtle updates in image processing methods while remaining specific enough to accept numerical variations within a reference version of the application. This result contributes to enhancing the reliability and reproducibility of data analyses by providing a robust and flexible method for stability testing.
\end{abstract}

\section{Introduction}

Data analyses can produce different results depending on the hardware and software conditions in which they are executed~\cite{gronenschild2012effects}, which has important repercussions in several disciplines. This paper investigates results stability tests whereby the outcome of data analysis is asserted to remain within acceptable variation bounds of a reference result. The primary challenge in developing such stability tests lies in the determination of acceptable bounds of variation around the reference result. We define such bounds from the numerical variability of the results, that is, from the variability inherent to numerical computations.

We focus on the use case of neuroimaging analyses, although our method applies to data analyses more broadly. For several decades, the neuroimaging community has developed advanced software tools (e.g., \emph{FSL}~\cite{jenkinson2012fsl}, \emph{FreeSurfer}~\cite{fischl2012freesurfer}, {\emph{ANTs}~\cite{avants2009advanced}, or \emph{AFNI}~\cite{COX1996162}) enabling researchers to study the human brain with unprecedented detail and precision. Since neuroimaging tools now underpin scientific findings in several disciplines, it is imperative to thoroughly test them. In particular, neuroimaging studies often follow subjects over multiple years, which requires data analyses to be consistent over substantial periods.

While existing research has focused on best practices for the accuracy and consistency of neuroimaging results~\cite{tustison2013instrumentation}, it is also important to consider the computational environments. Indeed, software packages or operating systems can significantly impact specific area studies such as brain surface reconstruction and cortical thickness quantification. Notably, research has highlighted substantial differences across software packages working on identical brain data~\cite{mikhael2019controlled}. Moreover, studies underscored the effects of operating systems on measured cortical thickness across various software packages and versions~\cite{glatard2015reproducibility}.

The present paper focuses primarily on the fMRIPrep software~\cite{esteban2019fmriprep}, a tool to pre-process Magnetic Resonance Imaging (MRI) data as a pre-requisite of any further analysis. fMRIPrep is an integrated data analysis pipeline for structural and functional MRI pre-processing. We focus on the pre-processing of structural MRI, which includes intensity non-uniformity correction, skull stripping, and spatial normalization to a brain template. In response to these computational challenges that impact result stability, fMRIPrep developers recently initiated long-term support (LTS) releases to guarantee results stability over multiple years which is critical in longitudinal analysis. This motivated the development of the stability tests presented in this paper.

Our stability tests leverage the numerical variability resulting that arises from using finite-precision arithmetic in calculations. Such variability primarily emerges due to changes in computational environments, which includes hardware architecture, parallelization scheme, operating system, and software dependencies. To estimate numerical variability, we rely on random rounding~\cite{forsythe1959reprint}, a stochastic arithmetic technique that randomly rounds floating-point operation results to the previous or to the next floating-point number. Stochastic arithmetic has been successfully applied to simulate numerical variability in various domains, including neuroimaging~\cite{salari2021accurate,kiar2021numerical}. Our approach is not specific to any particular numerical scheme and relies on a few statistical assumptions such as normality and independence, making it applicable to a wide range of scenarios.

In summary, the main contributions of this paper are the following:
\begin{itemize}
  \item Define a numerical variability approach to results stability tests;
  \item Build results stability tests for structural pre-processing in \fmriprep;
  \item Evaluate results of stability tests in several configurations.
\end{itemize}

\section{Results stability tests design}

Considering a data processing application $\Lambda$, the objective of our stability test is to determine whether the results generated by a different application $\tilde \Lambda$ significantly differ from the reference results produced by $\Lambda$. In practice, we are particularly interested in the case where $\tilde \Lambda$ corresponds to a different version of $\Lambda$ or the same version executed in a different execution environment (operating system, parallelization or hardware).

Given input data $I$, we assume that $\Lambda$ and $\tilde \Lambda$ produce images $X$ and $\tilde X$ sampled on the same imaging grid (i.e., they showcase the same number $v$ of voxels, and their orientation and resolution in physical coordinates do not differ by more than a very small error $\epsilon$).
We model $X$ as a random variable and we sample its distribution by computing $n$ random numerical perturbations of $\Lambda$, resulting in $n$ images $X_k$. Conversely, we compute $\tilde X$ without random perturbation using the IEEE-754 norm~\cite{ieee754}, to avoid computational overheads at test time.

The statistical test used for determining if '$\tilde X$ belongs to $X$' is discussed in detail in Subsection~\ref{subsec:statistical_model}. To sample reference distribution, we employ two methodologies, further elaborated in Subsection \ref{subsec:numerical_variablity_model}.
To capture anatomical variability and other sources of variability arising from the scanning device and sequence parameters, we test a diverse set of individuals collected from several studies, as discussed in Subsection \ref{subsec:data}.
To achieve precise reproducibility of the unperturbed sample, we control variables such as random seeding and multi-threading within the application, as detailed in Subsection \ref{subsec:computing_infrastructure}.
Table~\ref{tab:notations} summarizes our notations and Figure~\ref{fig:test_workflow} provides a comprehensive overview of our test workflow.

\subsection{Statistical model}
\label{subsec:statistical_model}

We preprocess the computed images $X,\tilde X$, leading to images $X^\perp, \tilde X^\perp$ (see~\ref{subsec:preprocessing}). To test if $\tilde X^\perp$ belongs to the distribution of $X^\perp$, we perform a $z$-test for each voxel within the union of $B_k$ masks, $\tilde x_i$ ($i\leq v$), using the mean $\hat \mu_i$ and the standard deviation $\hat \sigma_i$ estimated from the $k$ perturbed output.
The test computes a $p$-value $p_i$ under the null hypothesis $H_{0,i}$ that the tested voxel belongs to the reference distribution:
\begin{equation} \label{eq:pval}
  p_i(z_i) = 2 \left(1-\Phi(z_i)\right),
\end{equation}
where $\Phi$ is the cumulative distribution function of the normal centered Gaussian, and
\begin{equation*}
  z_i = \frac{\tilde x_i-\hat \mu_i}{\hat \sigma_i},
\end{equation*}
where is the intensity of a voxel $\tilde x_i \in \tilde X_k^\perp$, $v$ is the number of voxels within the union mask,
and $\hat \mu_i$ and $\hat \sigma_i$ are the mean and standard deviation voxel intensities estimated
from the $k$ perturbed results $X_k^\perp$.
$H_{0,i}$ is rejected when $p_i$ is lower than a threshold $\alpha$ that also defines the confidence level of the test (1-$\alpha$)\%.
The $z$-test assumes that perturbed voxel intensities are normally distributed.

\begin{figure}
  \centering
  \includegraphics[width=.95\columnwidth]{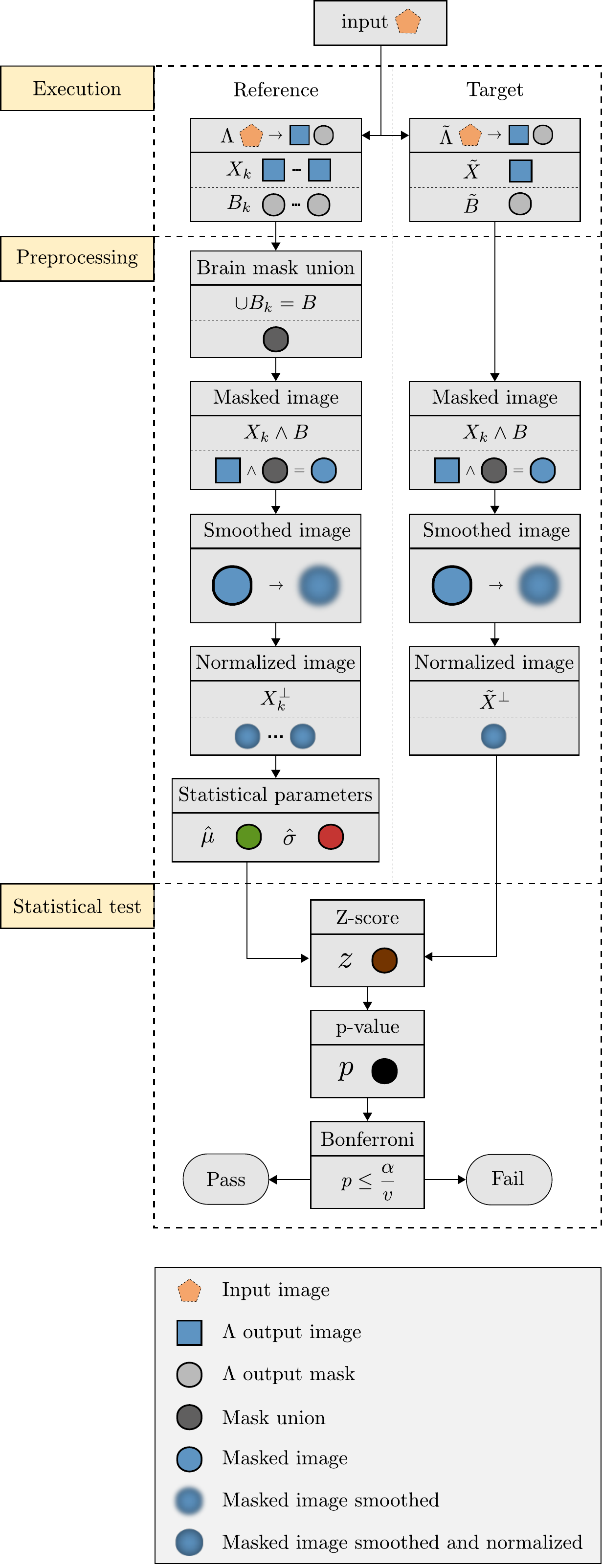}
  \caption{Stability test construction and evaluation. The reference application $\Lambda$ is executed $n$ times on the input data, with random perturbations, resulting in images $X_k$ from which a null distribution is built to test the result produced by target application $\tilde \Lambda$. All results undergo the same smoothing and normalization steps.}
  \label{fig:test_workflow}
\end{figure}

The test defined in Equation~\ref{eq:pval} consists of independent $z$-tests performed for each of the $v$ voxels
from each test image, resulting in a set of $v$ $p$-values $p_i$, $i \leq v$.
Because $v$ is in the order of one million (depending on the image's resolution and brain size), it is critical
to correct for multiple comparisons and set an upper bound to false positive tests~\cite{NICHOLS2007246}.
We adjust the significance $\alpha$ level with classical Bonferroni correction~\cite{farcomeni2008review}.
Therefore, the tested \fmriprep result is considered out of the reference distribution iif:
\begin{equation}
  \label{eq:bonferroni}
  \exists i \leq v \ \text{s.t.} \ p_i \leq \frac{\alpha}{v}.
\end{equation}

\begin{table}
  \centering
  \begin{tabular}{r|l}
    $\Lambda$          & reference application                                                \\
    $\tilde \Lambda$   & tested application                                                   \\
    $i$                & index of a voxel in B                                                \\
    $v$                & number of voxels in B                                                \\
    $k$                & index of a perturbed sample                                          \\
    $n$                & number of perturbed samples                                          \\
    $\tilde x_i$       & voxel intensity in image $\tilde X$                                  \\
    $z_i$              & z-score of voxel i                                                   \\
    $p_i$              & p-value of voxel i                                                   \\
    $\hat{\mu_i}$      & average voxel intensity across perturbed samples                     \\
    $\hat{\sigma_i}$   & standard deviation of voxel intensity across perturbed samples       \\
    $B_k$              & brain masks produced by random numerical perturbations of $\Lambda$  \\
    $B$                & union of $B_k$ masks                                                 \\
    $\tilde X$         & image produced by unperturbed $\tilde \Lambda$ using IEEE convention \\
    $X_k$              & images produced by random numerical perturbations of $\Lambda$       \\
    $X_k^{\perp}$      & $X_k$ masked with $B$, min-max scaled and spatially smoothed         \\
    $\hat{s}$          & average number of significant bits across the image                  \\
    $\mathcal{N}(0,1)$ & standard Gaussian distribution                                       \\
    $\Phi$             & cumulative distribution function of $\mathcal{N}(0,1)$               \\
    $\chi^2_{n-1}$     & Chi-2 distribution with n-1 degrees of freedom                       \\
  \end{tabular}
  \caption{Notations}
  \label{tab:notations}
\end{table}

\subsection{Numerical variability estimation}  
\label{subsec:numerical_variablity_model}

To estimate the distribution of reference results and compute $\hat \mu_i$ and $\hat \sigma_i$, we sample results distributions by applying two types of random numerical perturbations: (1) Random Rounding (RR), which randomly rounds function outputs in the \emph{GNU libmath} mathematical library, and (2) Random Seed (RS), which varies the random seed used in the application. In neuroimaging, random seeds are typically employed to initialize optimization processes encountered in spatial normalization.

Random Rounding (RR) consists in rounding the exact result of a floating-point arithmetic operation toward the previous or next floating-point number~\cite{forsythe1959reprint}. RR is equivalent to applying Monte-Carlo Arithmetic (MCA~\cite{parker1997monte}) to double-precision numbers with a virtual precision of 53 bits and to single-precision numbers with a virtual precision of 24 bits, which was shown to accurately simulate the effect of operating system updates on the structural MRI pre-processing pipelines of the Human Connectome Project (HCP) when applied to \emph{GNU libmath}~\cite{salari2021accurate}. Structural HCP pipelines consist of tools assembled from the \emph{FSL}~\cite{jenkinson2012fsl} and \emph{Freesurfer}~\cite{fischl2012freesurfer} toolboxes, which makes them conceptually very similar to the structural \fmriprep pipeline targeted by our study.

RR is rigorously implemented in several tools including \emph{CADNA}~\cite{jezequel2008cadna}, \emph{Verrou}~\cite{fevotte2016verrou}, and \emph{Verificarlo}~\cite{denis2016verificarlo}.
However, these tools incur substantial performance overheads which makes them hard to apply to compute-intensive applications. In addition, only \emph{Verrou} supports RR instrumentation of \emph{GNU libmath} as a standalone library~\cite{fevotte2019debugging}, and it does so by relying on quadruple precision, which is not scalable to entire neuroimaging pipelines.
Therefore, we implemented a fast, approximate RR method by randomly adding or removing 1 ulp (unit in the last place) to the outputs of \emph{GNU libmath}'s functions.
Our implementation, available on GitHub\footnote{\url{https://github.com/verificarlo/fuzzy/blob/master/docker/resources/libmath/fast/src/wrapping\_script.c}} under Apache 2.0 license,  only approximates RR as it applies a random perturbation to an already rounded result instead of to the exact result as done in rigorous implementations.
In practice, computing the exact result returned by \emph{GNU libmath}'s functions using tools like \emph{MPFR}~\cite{fousse2007mpfr} is too costly for our use case.

Random Seed (RS) and RR trigger different types of variability. RR can be applied transparently to any application while RS is more specific to the type of analysis. Conversely, RR incurs a substantial performance overhead whereas RS does not.

\subsection{Preprocessing of \fmriprep's outputs}
\label{subsec:preprocessing}

For the \fmriprep application, the main structural derivative produced is $X_k$, that is, the T1-weighted MRI image corrected for intensity non-uniformity using \texttt{N4BiasFieldCorrection} from \emph{ANTS} and transformed to template space using \texttt{antsRegistration}.
This file is named \texttt{desc-preproc\_T1w} in the \fmriprep outputs. In addition to $X_k$, \fmriprep produces a brain mask $B_k$, a segmentation into grey matter, white matter and cerebrospinal fluid tissues, as well as probability maps for each of these tissues.

Before computing the p-values in Equation~\ref{eq:pval}, we apply brain masking, smoothing, and intensity normalization to $X_k$. For brain masking, we mask $X_k$ with the union of the brain masks produced across all perturbed results. We use the union of the brain masks rather than their intersection to capture variability across $B_k$ masks.
For smoothing, we apply a spatial 3D Gaussian smoothing kernel with full-width at half-maximum (\fwhm) ranging from 0\,mm to 20\,mm.
The intensity values of $X_k$ are normalized to the [0, 1] range with min-max scaling.
The resulting preprocessed image $X_k^\perp$ is used as input for the stability test.

\subsection{Numerical stability measure}
\label{subsec:sigbits}

As a by-product of test construction, we can measure numerical variability in the application results, which provides valuable information about their numerical quality.
We quantify numerical stability for each voxel using the number of significant bits as a metric. This metric helps distinguish the bits containing signal from those containing only noise. We compute the number of significant bits $\hat{s}$ with probability $p_s=0.95$ and confidence $1-\alpha_s=0.95$ using the \emph{Significant Digits} package\footnote{\url{https://github.com/verificarlo/significantdigits}} (version 0.1.2).
\emph{Significant Digits} implements the Centered Normality Hypothesis approach described in~\cite{sohier2021confidence}:
\[
  \hat{s_i} = -\log_2 \left| \frac{\hat{\sigma_i}}{\hat{\mu_i}} \right| - \delta(n, \alpha_s, p_s),
\]

where $\hat{\sigma_i}$ and $\hat{\mu_i}$ are the voxelwise average and standard deviation over the $X_k^\perp$ perturbed results ($k \leq n$), and
\begin{equation}
  \delta(n, \alpha_s, p_s) = \log_2 \left( \sqrt{\frac{n-1}{\chi^2_{1-\alpha_s/2}}} \Phi^{-1} \left( \frac{p_s+1}{2} \right) \right)
\end{equation}
is a penalty term for estimating $\hat{s_i}$ with probability $p_s$ and confidence level $1-\alpha_s$ for a sample size $n$.
$\Phi^{-1}$ is the inverse cumulative distribution of the standard normal distribution and $\chi^2$ is the Chi-2 distribution with $n$-1 degrees of freedom.

\subsection{Data}
\label{subsec:data}

We selected eight test subjects from sub-datasets in the OpenNeuro~\cite{markiewicz2021openneuro} data-sharing platform, representing a diversity of ages, sex, and study designs. The datasets include a motion study with children (ds000256), a long-term memory study with young adults (ds001748), and a motor process study with adults (ds002338). In addition, two sub-datasets involve steps of the pipeline that can affect its reproducibility, namely different field maps (ds001600) and non-structural images (ds001771). Table~\ref{table:dataset_info} lists the dimension, voxels resolution, age and sex of each subject in the dataset.

\begin{table*}
  \begin{center}
    \begin{tabular}{c|c|l|c|c|c|c|c}
      Index & Dataset  & Subject     & Dimension ($x,y,z$)         & Voxel resolution            & Data type & Age     & Sex \\
            &          &             &                             & $mm^3$ ($x,y,z$)            &           & (years) &     \\
      \hline
      1     & ds000256 & sub-CTS201  & $256 \times 256 \times 256$ & $1.0 \times 1.0 \times 1.0$ & int16     & 8.68    & M   \\
      2     & ds000256 & sub-CTS210  & $224 \times 256 \times 256$ & $0.8 \times 0.8 \times 0.8$ & int16     & 7.63    & F   \\
      3     & ds001600 & sub-1       & $176 \times 256 \times 256$ & $1.0 \times 1.0 \times 1.0$ & int16     & -       & -   \\
      4     & ds001748 & sub-adult15 & $176 \times 240 \times 256$ & $1.0 \times 1.0 \times 1.0$ & float32   & 21      & M   \\
      5     & ds001748 & sub-adult16 & $176 \times 240 \times 256$ & $1.0 \times 1.0 \times 1.0$ & float32   & 21      & F   \\
      6     & ds001771 & sub-36      & $256 \times 320 \times 320$ & $0.8 \times 0.8 \times 0.8$ & int16     & 22      & F   \\
      7     & ds002338 & sub-xp201   & $176 \times 512 \times 512$ & $1.0 \times 0.5 \times 0.5$ & int16     & 41      & F   \\
      8     & ds002338 & sub-xp207   & $176 \times 512 \times 512$ & $1.0 \times 0.5 \times 0.5$ & int16     & 39      & M   \\
    \end{tabular}
  \end{center}
  \caption{Dimension, voxels resolutions, age and sex of each subject in the dataset.}
  \label{table:dataset_info}
\end{table*}

\subsection{Computing infrastructure}
\label{subsec:computing_infrastructure}

We processed the dataset using the Narval cluster managed by Calcul Qu\'ebec and part of the Digital Research Alliance of Canada.
With our job submission parameters, we could access 1,145 computing nodes with 64 cores per node and 2 $\times$ AMD Rome 7532 @ 2.40 GHz 256M cache L3. We executed \fmriprep in a Singularity container built from a Docker image available on DockerHub \texttt{yohanchatelain/fmriprep-fuzzy:20.2.1}.
The container image used \emph{Ubuntu} version 16.04.6 (LTS) with Linux kernel 4.18.0\footnote{Kernel build 4.18.0-372.19.1.el8\_6.x86\_64}, \emph{GNU libc/libmath} 2.23, and \fmriprep 20.2.1.
We used \emph{Fuzzy} \texttt{v0.9.1-a} built with \emph{Verificarlo} version \texttt{v0.9.1}.

Further spurious variability can source from the ever-changing initial point of all random number generators reliant on stochastic processes that are involved in the processing (e.g., optimization).
To control for such variability, \fmriprep's `master' random seed (\texttt{--random-seed=42}) as well as the skull stripping random seed were fixed. We also fixed multi-threading to perform calculations in a deterministic order.

\section{Results}

We computed $n$=30 perturbed results for each of the eight subjects and for the two types of perturbations RR and RS. For RS, we used \fmriprep's command-line interface to set the random seed in all the pipeline components. We also computed an unperturbed IEEE result for each subject using the random seed used in RR (42). For each type of perturbation, we measured the numerical stability of \fmriprep in terms of significant bits, and we built the stability test using different sizes of smoothing kernel and different confidence levels.

\subsection{Measured numerical variability was high and it varied across subjects}

\begin{figure}
  \centering
  \includegraphics[width=\linewidth]{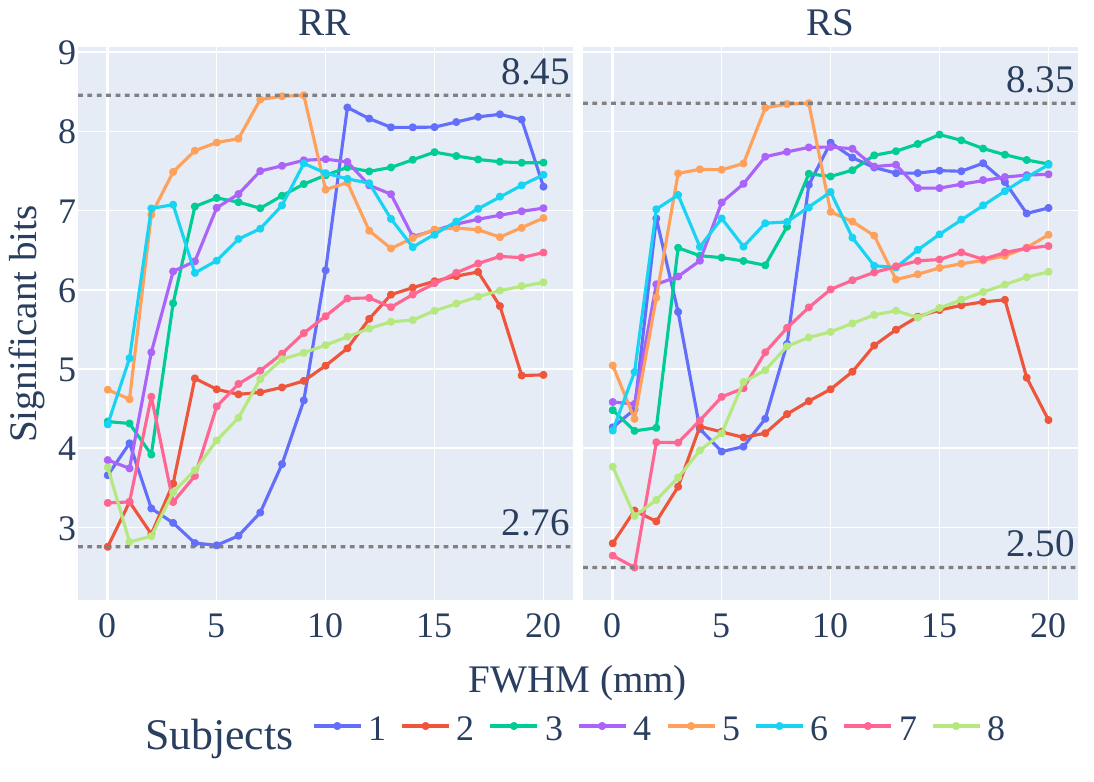}
  \caption{Voxel-wise means of significant bits
    measured across n=30 perturbed samples for RR and RS perturbations and 8
    subjects.}
  \label{fig:significant-digits}
\end{figure}
Overall, the two types of perturbations (RR and RS) resulted in numerical uncertainties of comparable magnitude and behavior (Figure~\ref{fig:significant-digits}), which supports the validity of our results. The measured numerical variability was high, with mean significant bits ranging from 2.5 bits to 8.5 bits out of the 12 bits available in the data\footnote{The voxel intensity is encoded on 12 bits although it is embedded in a 16 bits format.}. The application appears to be highly sensitive to numerical and random seed perturbations.

We noted substantial discrepancies in numerical stability across subjects. For a given smoothing kernel size, the number of significant bits frequently varied in the ratio of 1 to 3 across subjects. Overall, smoothing tended to reduce numerical variability, however, this behavior was in general not monotonous and impacted subjects differently. The observed between-subject variability
demonstrates the importance of evaluating such applications on a representative set of datasets.

The numerical variability measured across perturbed samples showed regional variations compatible with anatomical features (Figure~\ref{fig:uncertainty-maps} and Appendix~\ref{appendix:numerical_uncertainty}). In particular, variability was maximal at the border of the brain mask, and it was overall higher in the gray matter than in the white matter.
This is consistent with previous observations of numerical variability in structural brain image analysis~\cite{salari2021accurate}.
In addition, numerical variability was also maximal in some focal regions, suggesting that spatial normalization may be unstable in these regions.
Our stability test will be more permissive, resulting in fewer rejections of voxels in these regions.

\begin{figure*}
  \uncertaintyMapDiscreteAll{15}
  \vspace*{-20pt}\hspace{15pt}
  \includegraphics*[width=.6\linewidth]{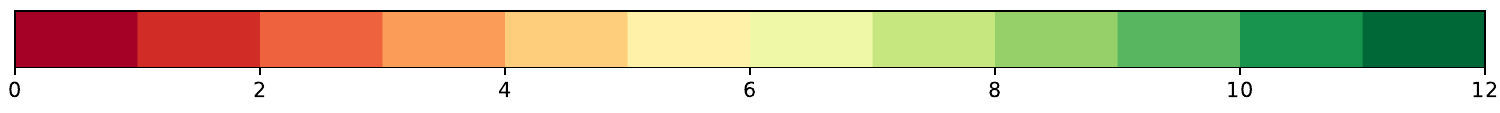} \\
  \vspace*{-5pt}
  significant bits
  \caption{Numerical variability measured for subjects 1 to 8 (from top to bottom) across n=30 perturbed samples, with FWHM\,=\,15\,mm. Uncertainty maps for other smoothing kernel sizes are available in Appendix~\ref{appendix:numerical_uncertainty}.}
  \label{fig:uncertainty-maps}
\end{figure*}

\subsection{The results stability test passed sanity checks but usable FWHM and \texorpdfstring{$\alpha$}{α} values were data-dependent}

We implemented three different sanity checks to evaluate the relevance of the stability test.
The leave-one-out check (\ref{subsec:loo_check}) evaluates the specificity of the stability test, that is, its ability to accept results produced by the reference application $\Lambda$.
The IEEE check (\ref{subsec:ieee_check}) evaluates the specificity of the test by checking that it accepts the unperturbed application result and its sensitivity through the rejection of results produced by a different subject than the original result.
The corrupted template check (\ref{subsec:template_check}) evaluates the sensitivity of the stability test, that is, its ability to reject results produced by a corrupted version of the reference application.

\subsubsection{Leave-one-out check}
\label{subsec:loo_check}
We implemented a ``leave-one-out" (LOO) evaluation by constructing the stability test $n$ times for $n-1$ perturbed results and applying it to the remaining perturbed result. By construction, the remaining perturbed result is sampled from the distribution of results produced by $\Lambda$ and should therefore be accepted by the stability test.
To define a clear passing criterion for the LOO check, we modeled the LOO check using a binomial variable $B(n,1-\alpha)$ where $n$ is the number of LOO iterations and $1-\alpha$ is the probability that a perturbed sample is accepted by the stability test. Under $H_0$ for all voxels, we expect the following bound to be verified:
\[
  1-F(\mathds{1}_n;n,1-\alpha) \leq \alpha_0
\]
where $F(x;n,p)$ is the cumulative distribution function of the Binomial law $B(n,p)$, and $\alpha_0=0.05$.

\paragraph*{Results} We applied the LOO check for different confidence values ($1-\alpha$) and different FWHM  values for the RR and RS perturbations (Figure~\ref{fig:loo_bonferroni}). As expected, the stability test became increasingly permissive for increasing values of $\alpha$ (reduced confidence) and increasing values of FWHM. As previously observed, RR and RS behaved similarly overall. For each subject, there were values of $\alpha$ and FWHM such that the LOO check passed, which demonstrates the good specificity of the stability test.

However, the values of $\alpha$ and FWHM for which the LOO check passed importantly varied across subjects, presumably due to between-subject variations in input data quality and instability modes of spatial normalization. For instance, to pass the LOO check with $\alpha=0.05$ for RR perturbations, subjects 5, 6, 7 and 8 required a smoothing size of FWHM\,=\,12\,mm and subjects 1 and 4 required FWHM\,=\,15\,mm. Subjects 2 and 3 required $\alpha=0.15$ to pass the LOO check. Such discrepancies are not surprising given the heterogeneity in numerical variability previously observed across subjects. In practice, different values of $\alpha$ and FWHM must be used for each subject in the test dataset.

\begin{figure}
  \centering
  \includegraphics[width=\linewidth]{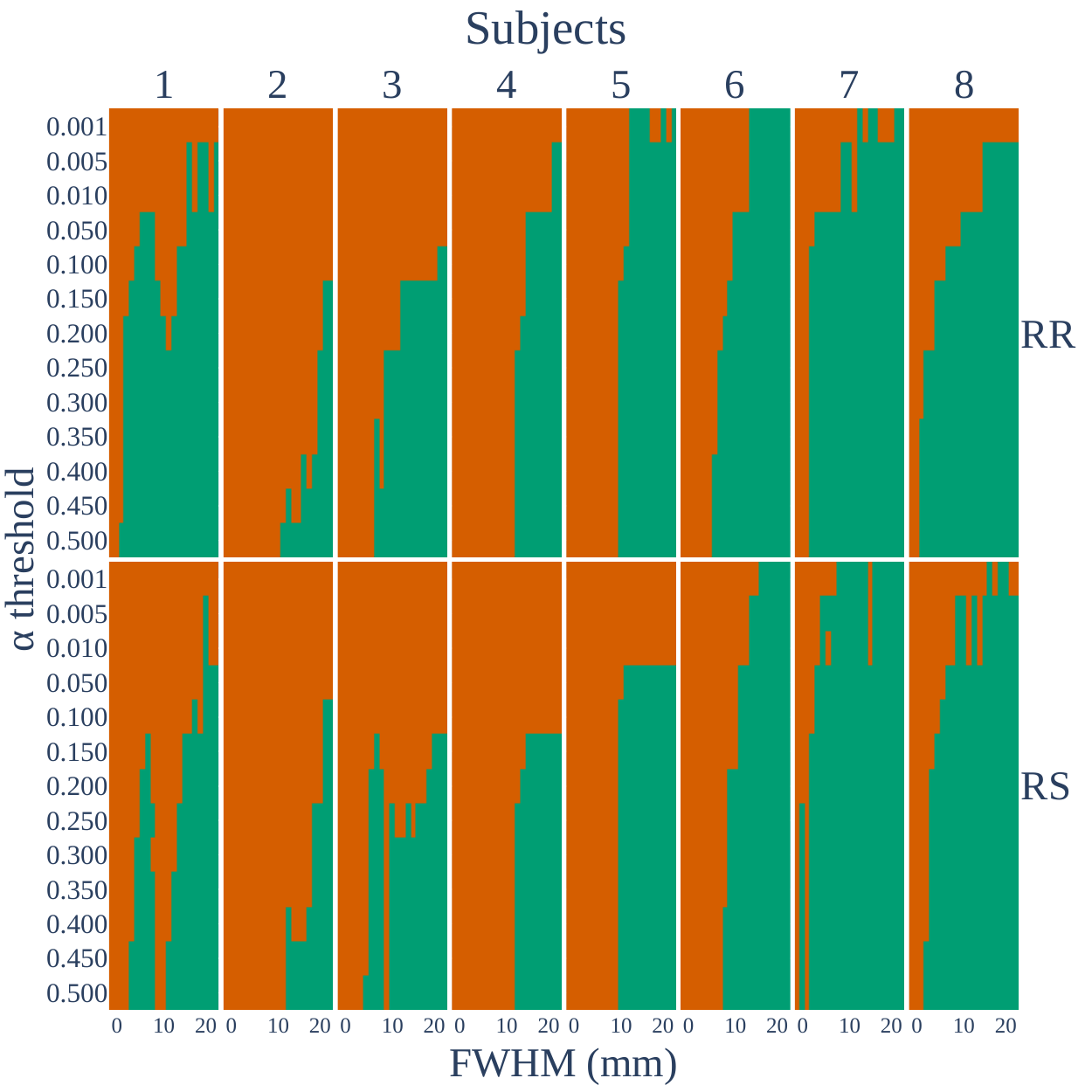}
  \caption{Leave-one-out evaluation of stability test for subjects 1 to 8.
    Red/green: rejected/accepted by a binomial one-tailed test with sample size $n=30$ and confidence level $1-\alpha_0=0.95$.}
  \label{fig:loo_bonferroni}
\end{figure}

\subsubsection{IEEE check}
\label{subsec:ieee_check}

We constructed the stability test from the $n$ perturbed results for each subject and applied it to the IEEE results (one per subject). The purpose of this check was twofold: (within subjects) to verify that IEEE results passed the stability test built from the reference distribution of their corresponding subject and (between subjects) to verify that IEEE results failed the stability test built from the reference distribution of other subjects.

\paragraph*{Within subjects} for each subject, there was an $\alpha$ and FWHM pair such that all the within-subject IEEE checks passed (Figure~\ref{fig:ieee-check}). In particular, the within-subject IEEE check passed with FWHM\,=\,15\,mm for all subjects and $\alpha$ values for RR perturbations. For low FWHM sizes, the stability test rejected the IEEE sample of the reference subject, suggesting a lack of specificity in such cases.

\paragraph*{Between subjects} The stability tests successfully rejected all the IEEE samples coming from other subjects, for all combinations of $\alpha$ and FWHM values, and consistently for both RR and RS perturbations. Therefore the stability test is sufficiently sensitive to detect between-subject variability even with a high smoothing kernel size.

We conclude from this experiment that the stability test is sensitive enough to reject results obtained from different subjects using the reference application and accept results obtained from the same subject using the reference application executed with random perturbations.

\begin{figure}
  \centering
  \includegraphics[width=\linewidth]{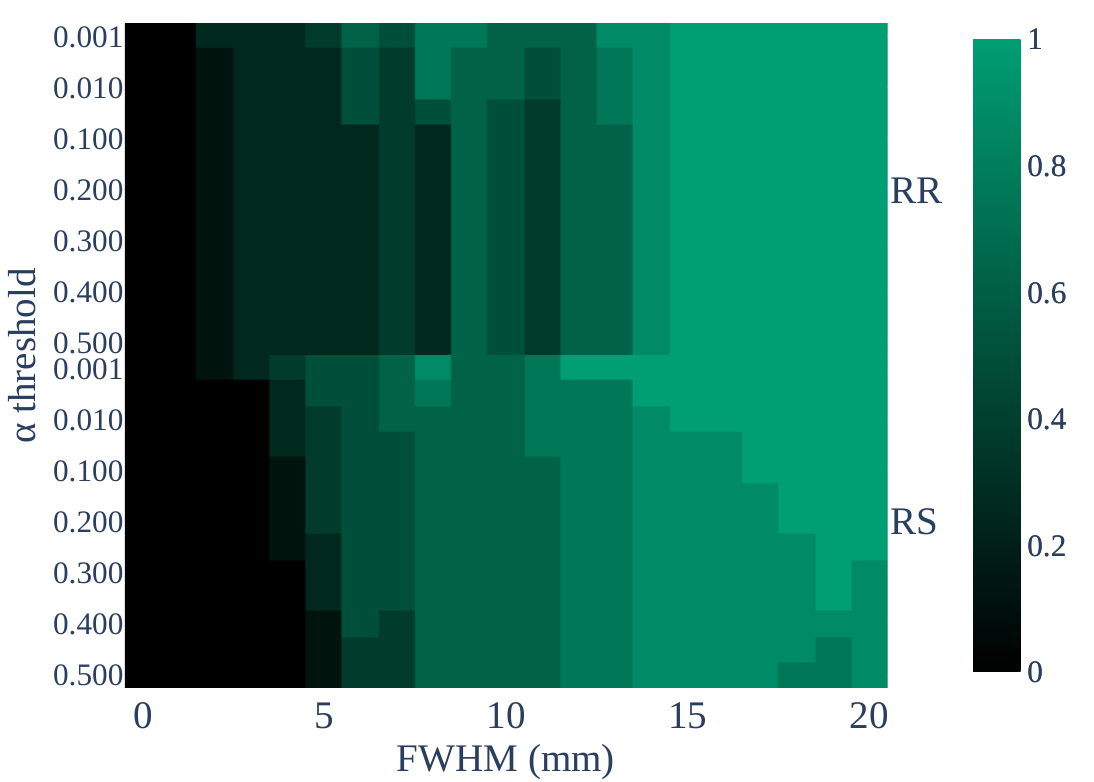}
  \caption{Ratios of successful stability tests for the within-subject IEEE check. The between-subject IEEE check (not displayed in the figure) successfully rejected data computed from different subjects for all $\alpha$ and FWHM combinations, demonstrating excellent sensitivity.
  }
  \label{fig:ieee-check}
\end{figure}

\subsubsection{Corrupted template check}
\label{subsec:template_check}

Multiple brain templates exist to spatially normalize subject data to a common space, and template selection substantially impacts the results~\cite{li2021moving}. Hence, errors in the template should lead to substantial differences in the results. The purpose of this check was to verify that results obtained from corrupted templates were correctly rejected by the stability test.
To do so, we generated corrupted versions of the MNI152NLin2009cAsym template used by \fmriprep where we incrementally zeroed the intensity of an increasing fraction of brain voxels selected uniformly. We then executed \fmriprep in IEEE mode (without random perturbations) for each resulting corrupted template and each subject.

\paragraph*{Results} The stability test correctly rejected results obtained with the corrupted template above a subject-dependent threshold of corrupted voxels (Figure~\ref{fig:template_bonferroni}). Smoothing had a measurable effect that did not manifest clearly across subjects. We conclude that the stability test is sensitive to changes in the brain template which is a common source of variability in neuroimaging.

\begin{figure}
  \centering
  \includegraphics[width=\linewidth]{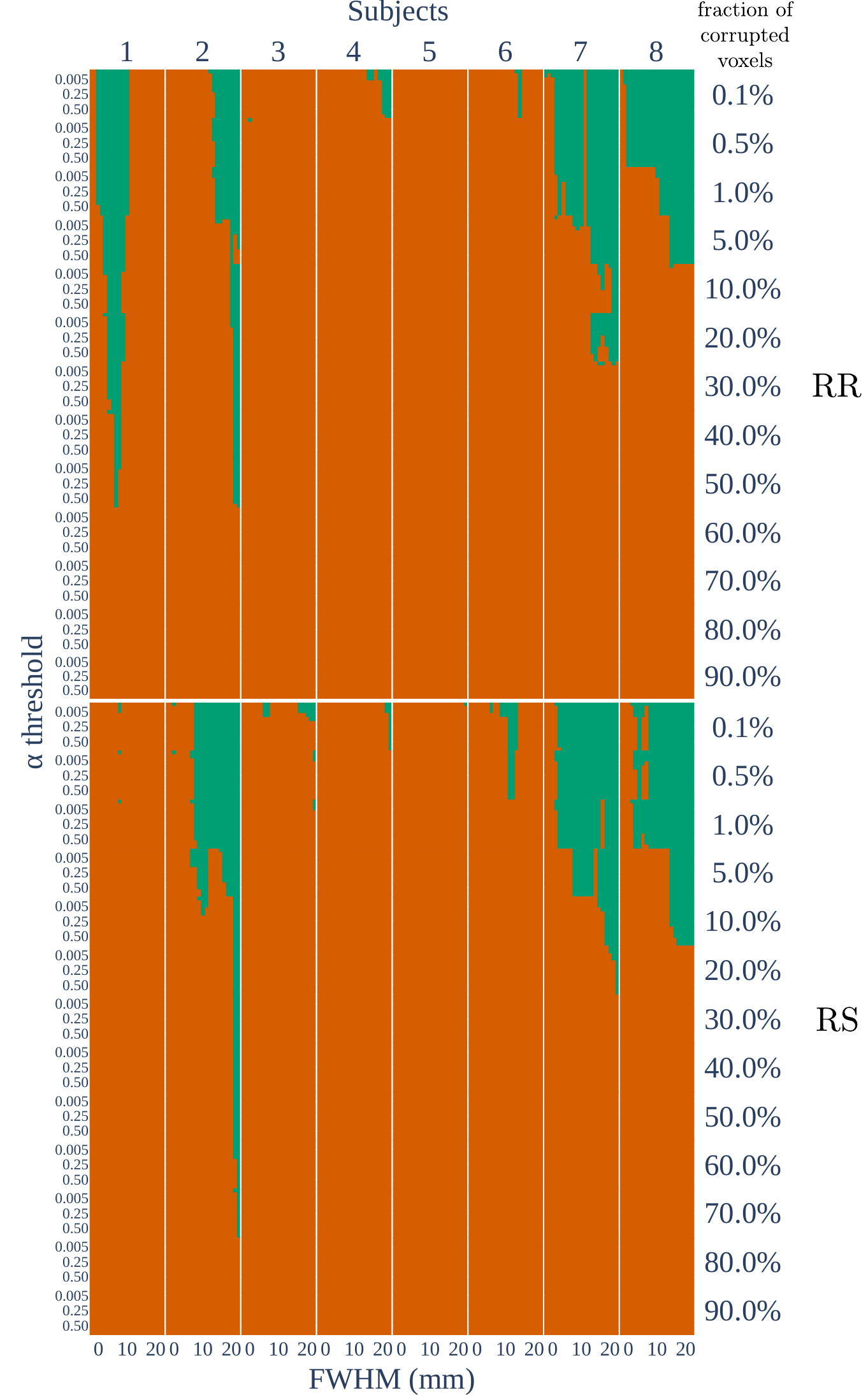}
  \caption{Corrupted template check for RR (top) and RS (bottom) modes. The stability tests correctly rejected results produced with corrupted templates beyond a subject-dependent threshold of corrupted voxels.}
  \label{fig:template_bonferroni}
\end{figure}

\subsection{The results stability test detected subtle updates in \fmriprep image processing methods}

The previous sanity checks demonstrated a satisfying level of sensitivity and specificity of our stability test for simple scenarios. Subsequently, we applied it to our main motivating use case: the detection of results differences in \fmriprep LTS versions. We tested the results produced by different versions of \fmriprep, having built the results stability test for version 20.2.1. The goal of this experiment was to evaluate the ability of the stability test to detect significant software updates in \fmriprep. The stability test accepted results produced by versions 20.2.0 to 20.2.4 for all subjects (Figure~\ref{fig:version_bonferroni}). However, the test rejected results produced by versions 20.2.5 and onwards, suggesting a substantial change in the analysis methods from version 20.2.5. Investigations with the \fmriprep developers revealed that the interpolation scheme involved in the resampling of piece-wise-smooth brain segmentation maps from \emph{FreeSurfer}'s preferred format (MGZ) format into NIfTI 1.0 format was changed in version {20.2.5} from trilinear interpolation to nearest neighbor (see pull-request \href{https://github.com/nipreps/smriprep/pull/268}{\url{github.com/nipreps/smriprep/pull/268}}). The ability of our results stability test to detect such subtle changes in the analysis supports its relevance in the release process of long-term support data analysis software.

\begin{figure}
  \centering
  \includegraphics[width=\linewidth]{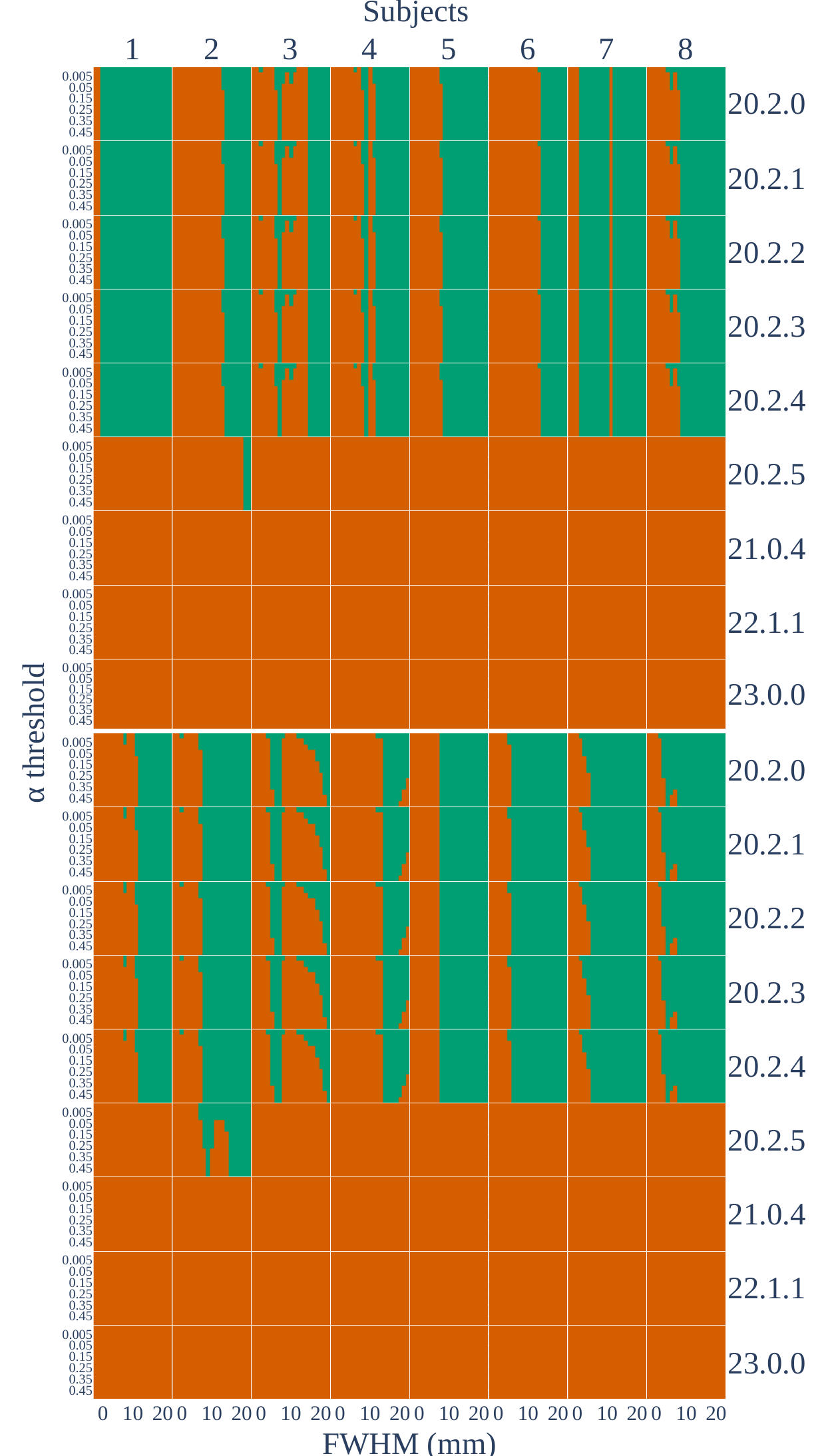}
  \caption{Application of the stability test across \fmriprep versions, with the reference distribution constructed using version 20.2.1. The test indicated a divergence in the results beginning with version 20.2.5. Investigations showed a substantial change in the \fmriprep analysis methods since version 20.2.5, correctly detected by the stability test.}
  \label{fig:version_bonferroni}
\end{figure}

\section{Discussion}


We introduced a novel approach to build results stability tests for data analyses using numerical variability, with a specific focus on neuroimaging and the widely used \fmriprep software. While libraries to test numerical stability exist and have been applied in other contexts, to our knowledge, this is one of the first attempts to systematically apply such an approach in the domain of neuroimaging and possibly in other data analysis domains.
This approach does not require an exact solution, a reference solution, or even an acceptable bound of variation around the computed solution.
Applying this approach to \fmriprep, a highly popular neuroimaging preprocessing pipeline, we demonstrated non-negligible, subject-specific numerical errors that the tool introduces in the results as a consequence of numerical instability.
In addition, we showed how the framework successfully detected subtle algorithmic updates through \fmriprep releases, while remaining specific enough to accept numerical variations within a reference version of the application.
Based on our findings, we contend that benchmarking neuroimaging and other computational tools is essential. This not only includes accuracy against gold or \emph{golden} standards in cases where a ground-truth reference is absent and performance but also numerical stability.

Our stability test represents a practical tool that has the potential for widespread application, not just limited to the \fmriprep development environment. Our results stability test is compatible with continuous integration platforms, enabling automated validation of new commits. By choosing suitable data and confidence values, our results stability test can serve as a valuable addition to conventional code testing methodologies. Additionally, the figure templates presented in this paper can be reused to create informative visualizations for dashboards or similar data presentation platforms. This way, the results are easily interpretable and actionable.

Designing the results stability test involved multiple choices that we discuss hereafter.
First, we employed a parametric statistical test (Equation~\ref{eq:pval}) that is less compute-intensive than non-parametric methods that typically demand larger sample sizes --- in our case, more than $100$ repetitions per subject --- to achieve acceptable confidence levels ($\geq 0.95$).
Indeed, the computational and storage costs required to apply stochastic arithmetic to neuroimaging tools such as \fmriprep are high, although they are incurred only during the construction of the test and not during its evaluation, which therefore does not impact test users. Moreover, our stability test operates at the level of individual voxels and is accepted or rejected globally for a given image. Refining the test by regions of interest in the image could provide additional insights about the application behavior.
The Bonferroni correction for multiple comparisons is conservative in our case given that the voxel intensities in a brain image are not independent.
Possible alternatives to this correction are presented in~\cite{NICHOLS2007246}. The use of Bonferroni correction reduces the sensitivity of our test and increases its specificity.

The random rounding method adopted to sample the null distribution of results produced by the reference application can be applied regardless of the type of input and output data involved. In our \fmriprep use case, random rounding resulted in comparable variability to the one obtained from varying the random seed, which increased the confidence in our results. However, it is important to note that random rounding and random seed are different types of perturbations, and there is no guarantee of producing comparable variability patterns in general. We applied random rounding to functions of the \emph{GNU libmath} library since previous experiments had shown that neuroimaging analyses commonly rely on these functions. Other types of applications may require that random rounding be applied to other libraries. For instance, the work in~\cite{pepe2022numerical} reports on the use of random rounding for deep neural networks, which required instrumenting the entire Tensorflow application and its dependencies.

The size of the spatial smoothing kernel had a notable impact on the performance of the stability test. Smoothing was required for the test to accept results produced by the reference application and therefore reach an acceptable level of specificity.
The smoothing process transforms the distribution of voxel intensities towards a Gaussian distribution, thus better aligning with the main assumption of the test.
Smoothing is a prevalent operation in neuroimage analysis to improve the signal-to-noise ratio. However, FWHM values usually remain lower than 8\,mm while our test required 15\,mm for most subjects. It would be interesting to further investigate why such large smoothing kernel sizes were required since large smoothing kernel sizes reduce the sensitivity of our test.

The results stability test behaved quite differently across the 8 test subjects, which motivates the inclusion of various datasets in such test cases.
To build our test, we selected subjects to reflect diverse acquisition parameters, resolutions, ages, and sexes.
However, it would be interesting to investigate two distinct effects: (1) differences between subjects, and (2) differences due to scanners or sequence parameters. For the first case, using a dataset from a single study is more appropriate since it consists of subjects who have undergone identical experimental protocols and procedures, thereby ensuring better control over acquisition parameters.
In the second case, to examine the impact of differing scanners or sequence parameters, a traveling subject approach would be more effective. In this method, a single subject undergoes MRI scans at various locations, using different scanners or imaging protocols. This controlled approach provides valuable insight into the influence of these variables

In future work, we plan to extend our methodology to other data types and investigate its applicability under different statistical hypotheses.
Results stability tests could be applied to diverse scientific domains beyond neuroimaging, further enhancing the reliability and reproducibility of computational results. In the short term, we plan to extend our test to functional neuroimaging data, which involves 4D images.

\section*{Acknowledgments}

Computations were made on the Narval supercomputer from \'Ecole de Technologie
Sup\'erieure (ETS, Montr\'eal), managed by Calcul Québec and the Digital Research Alliance of Canada. The
operation of this supercomputer is funded by the Canada Foundation for
Innovation (CFI), Ministère de l’Économie, des Sciences et de l’Innovation du
Québec (MESI) and le Fonds de recherche du Québec – Nature et technologies
(FRQ-NT).

\bibliographystyle{IEEEtran}
\bibliography{main}

\begin{thebibliography}{10}
\providecommand{\url}[1]{#1}
\csname url@rmstyle\endcsname
\providecommand{\newblock}{\relax}
\providecommand{\bibinfo}[2]{#2}
\providecommand\BIBentrySTDinterwordspacing{\spaceskip=0pt\relax}
\providecommand\BIBentryALTinterwordstretchfactor{4}
\providecommand\BIBentryALTinterwordspacing{\spaceskip=\fontdimen2\font plus
\BIBentryALTinterwordstretchfactor\fontdimen3\font minus
  \fontdimen4\font\relax}
\providecommand\BIBforeignlanguage[2]{{%
\expandafter\ifx\csname l@#1\endcsname\relax
\typeout{** WARNING: IEEEtran.bst: No hyphenation pattern has been}%
\typeout{** loaded for the language `#1'. Using the pattern for}%
\typeout{** the default language instead.}%
\else
\language=\csname l@#1\endcsname
\fi
#2}}

\bibitem{gronenschild2012effects}
E.~H. Gronenschild, P.~Habets, H.~I. Jacobs, R.~Mengelers, N.~Rozendaal,
  J.~Van~Os, and M.~Marcelis, ``The effects of freesurfer version, workstation
  type, and macintosh operating system version on anatomical volume and
  cortical thickness measurements,'' \emph{PloS one}, vol.~7, no.~6, p. e38234,
  2012.

\bibitem{jenkinson2012fsl}
M.~Jenkinson, C.~F. Beckmann, T.~E. Behrens, M.~W. Woolrich, and S.~M. Smith,
  ``Fsl,'' \emph{Neuroimage}, vol.~62, no.~2, pp. 782--790, 2012.

\bibitem{fischl2012freesurfer}
B.~Fischl, ``Freesurfer,'' \emph{Neuroimage}, vol.~62, no.~2, pp. 774--781,
  2012.

\bibitem{avants2009advanced}
B.~B. Avants, N.~Tustison, G.~Song, \emph{et~al.}, ``Advanced normalization
  tools (ants),'' \emph{Insight j}, vol.~2, no. 365, pp. 1--35, 2009.

\bibitem{COX1996162}
\BIBentryALTinterwordspacing
R.~W. Cox, ``Afni: Software for analysis and visualization of functional
  magnetic resonance neuroimages,'' \emph{Computers and Biomedical Research},
  vol.~29, no.~3, pp. 162--173, 1996.
\BIBentrySTDinterwordspacing

\bibitem{tustison2013instrumentation}
N.~J. Tustison, H.~J. Johnson, T.~Rohlfing, A.~Klein, S.~S. Ghosh, L.~Ibanez,
  and B.~B. Avants, ``Instrumentation bias in the use and evaluation of
  scientific software: recommendations for reproducible practices in the
  computational sciences,'' p. 162, 2013.

\bibitem{mikhael2019controlled}
S.~S. Mikhael and C.~Pernet, ``A controlled comparison of thickness, volume and
  surface areas from multiple cortical parcellation packages,'' \emph{BMC
  bioinformatics}, vol.~20, no.~1, pp. 1--12, 2019.

\bibitem{glatard2015reproducibility}
T.~Glatard, L.~B. Lewis, R.~Ferreira~da Silva, R.~Adalat, N.~Beck, C.~Lepage,
  P.~Rioux, M.-E. Rousseau, T.~Sherif, E.~Deelman, \emph{et~al.},
  ``Reproducibility of neuroimaging analyses across operating systems,''
  \emph{Frontiers in neuroinformatics}, vol.~9, p.~12, 2015.

\bibitem{esteban2019fmriprep}
O.~Esteban, C.~J. Markiewicz, R.~W. Blair, C.~A. Moodie, A.~I. Isik,
  A.~Erramuzpe, J.~D. Kent, M.~Goncalves, E.~DuPre, M.~Snyder, \emph{et~al.},
  ``fmriprep: a robust preprocessing pipeline for functional mri,''
  \emph{Nature methods}, vol.~16, no.~1, pp. 111--116, 2019.

\bibitem{forsythe1959reprint}
G.~E. Forsythe, ``Reprint of a note on rounding-off errors,'' \emph{SIAM
  review}, vol.~1, no.~1, p.~66, 1959.

\bibitem{salari2021accurate}
A.~Salari, Y.~Chatelain, G.~Kiar, and T.~Glatard, ``Accurate simulation of
  operating system updates in neuroimaging using monte-carlo arithmetic,'' in
  \emph{Uncertainty for Safe Utilization of Machine Learning in Medical
  Imaging, and Perinatal Imaging, Placental and Preterm Image Analysis}.\hskip
  1em plus 0.5em minus 0.4em\relax Cham: Springer International Publishing,
  2021, pp. 14--23.

\bibitem{kiar2021numerical}
G.~Kiar, Y.~Chatelain, P.~de~Oliveira~Castro, E.~Petit, A.~Rokem, G.~Varoquaux,
  B.~Misic, A.~C. Evans, and T.~Glatard, ``Numerical uncertainty in analytical
  pipelines lead to impactful variability in brain networks,'' \emph{PloS one},
  vol.~16, no.~11, p. e0250755, 2021.

\bibitem{ieee754}
M.~S. Committee, ``{IEEE} standard for floating-point arithmetic,'' \emph{IEEE
  Std 754-2019 (Revision of IEEE 754-2008)}, pp. 1--84, 2019.

\bibitem{NICHOLS2007246}
\BIBentryALTinterwordspacing
T.~Nichols, ``Chapter 20 - false discovery rate procedures,'' in
  \emph{Statistical Parametric Mapping}, K.~Friston, J.~Ashburner, S.~Kiebel,
  T.~Nichols, and W.~Penny, Eds.\hskip 1em plus 0.5em minus 0.4em\relax London:
  Academic Press, 2007, pp. 246--252.
\BIBentrySTDinterwordspacing

\bibitem{farcomeni2008review}
A.~Farcomeni, ``A review of modern multiple hypothesis testing, with particular
  attention to the false discovery proportion,'' \emph{Statistical methods in
  medical research}, vol.~17, no.~4, pp. 347--388, 2008.

\bibitem{parker1997monte}
D.~S. Parker, \emph{Monte Carlo arithmetic: exploiting randomness in
  floating-point arithmetic}.\hskip 1em plus 0.5em minus 0.4em\relax Citeseer,
  1997.

\bibitem{jezequel2008cadna}
F.~J{\'e}z{\'e}quel and J.-M. Chesneaux, ``Cadna: a library for estimating
  round-off error propagation,'' \emph{Computer Physics Communications}, vol.
  178, no.~12, pp. 933--955, 2008.

\bibitem{fevotte2016verrou}
F.~F{\'e}votte and B.~Lathuiliere, ``{VERROU}: a {CESTAC} evaluation without
  recompilation,'' \emph{SCAN 2016}, p.~47, 2016.

\bibitem{denis2016verificarlo}
C.~Denis, P.~de~Oliveira~Castro, and E.~Petit, ``Verificarlo: checking floating
  point accuracy through monte carlo arithmetic,'' in \emph{2016 IEEE 23nd
  Symposium on Computer Arithmetic (ARITH)}, 2016.

\bibitem{fevotte2019debugging}
F.~F{\'e}votte and B.~Lathuiliere, ``Debugging and optimization of hpc programs
  with the verrou tool,'' in \emph{2019 IEEE/ACM 3rd International Workshop on
  Software Correctness for HPC Applications (Correctness)}.\hskip 1em plus
  0.5em minus 0.4em\relax IEEE, 2019, pp. 1--10.

\bibitem{fousse2007mpfr}
L.~Fousse, G.~Hanrot, V.~Lef{\`e}vre, P.~P{\'e}lissier, and P.~Zimmermann,
  ``Mpfr: A multiple-precision binary floating-point library with correct
  rounding,'' \emph{ACM Transactions on Mathematical Software (TOMS)}, vol.~33,
  no.~2, pp. 13--es, 2007.

\bibitem{sohier2021confidence}
D.~Sohier, P.~D.~O. Castro, F.~F{\'e}votte, B.~Lathuili{\`e}re, E.~Petit, and
  O.~Jamond, ``{Confidence Intervals for Stochastic Arithmetic},'' \emph{ACM
  Transactions on Mathematical Software (TOMS)}, vol.~47, no.~2, pp. 1--33,
  2021.

\bibitem{markiewicz2021openneuro}
C.~J. Markiewicz, K.~J. Gorgolewski, F.~Feingold, R.~Blair, Y.~O. Halchenko,
  E.~Miller, N.~Hardcastle, J.~Wexler, O.~Esteban, M.~Goncavles, \emph{et~al.},
  ``The openneuro resource for sharing of neuroscience data,'' \emph{Elife},
  vol.~10, p. e71774, 2021.

\bibitem{li2021moving}
X.~Li, L.~Ai, S.~Giavasis, H.~Jin, E.~Feczko, T.~Xu, J.~Clucas, A.~Franco,
  A.~S{\'o}lon~Heinsfeld, A.~Adebimpe, \emph{et~al.}, ``Moving beyond
  processing and analysis-related variation in neuroscience,'' \emph{BioRxiv},
  pp. 2021--12, 2021.

\bibitem{pepe2022numerical}
I.~G. Pepe, Y.~Chatelain, G.~Kiar, and T.~Glatard, ``Numerical stability of
  deepgoplus inference,'' \emph{arXiv preprint arXiv:2212.06361}, 2022.

\end{thebibliography}

\newpage

\section{Biography Section}

\vspace{11pt}


\begin{IEEEbiography}[{\includegraphics[width=1in,height=1.25in,clip,keepaspectratio]{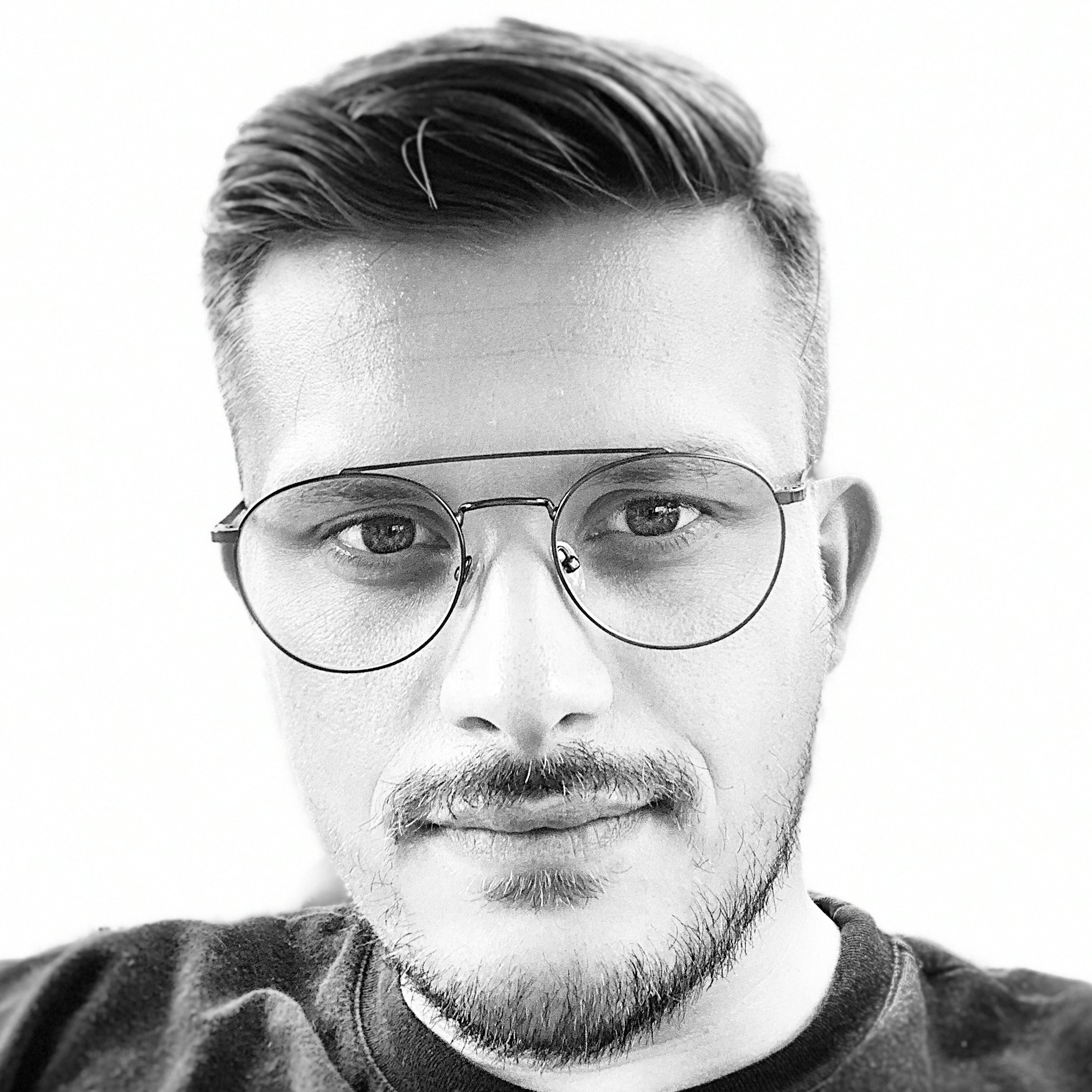}}]{Yohan Chatelain}
  is a postdoctoral researcher in the
  Big Data Infrastructures for Neuroinformatics lab at the University of
  Concordia, Montreal, Canada.
  He received his Ph.D. degree from the University of Paris-Saclay
  (UVSQ), Versailles, France in 2019. Yohan research topics include computer
  arithmetic, high-performance computing, and reproducibility. Yohan aims at
  democratizing the use of the analysis of stability for scientific computing
  codes through automatic tools to improve numerical quality.
\end{IEEEbiography}

\begin{IEEEbiography}[{\includegraphics[width=1in,height=1.25in,clip,keepaspectratio]{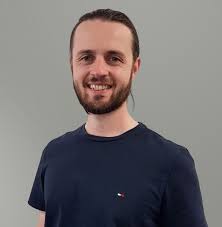}}]{Loïc Tetrel}
  is an R\&D Engineer on Kitware Europe’s computer vision team.
  Prior to joining Kitware, Loïc was a Data Scientist in Neuroscience at SIMEXP lab (University of Montreal) where he worked on neuroimaging software, machine learning, and research data platforms. He contributed to TensorFlow, Nilearn and Binderhub. Loïc also worked at Straumann Group as a computer vision engineer for 2 years, mainly working on 3D reconstruction algorithms.
  He holds an engineering degree from INSA Lyon and an M.A.Sc from ETS Montréal.
  At Kitware Europe, Loïc brings his competencies in Machine Learning and 3D modeling to the team, in particular in the medical domain.
\end{IEEEbiography}

\begin{IEEEbiography}[{\includegraphics[width=1in,height=1.25in,clip,keepaspectratio]{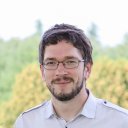}}]{Christopher J. Markiewicz}
  is a software developer for the Poldrack Lab at Stanford University and a research affiliate of the McGovern Institute for Brain Research. He has been a core developer of fMRIPrep and FitLins, and is a contributor to and maintainer of several open-source, Python neuroimaging libraries, including Nipype and NiBabel.
\end{IEEEbiography}

\begin{IEEEbiography}[{\includegraphics[width=1in,height=1.25in,clip,keepaspectratio]{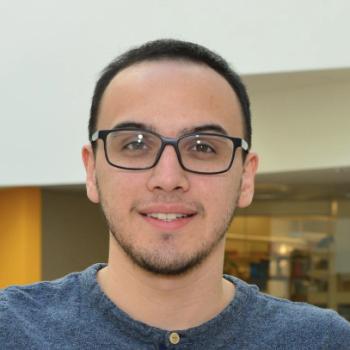}}]{Mathias Goncalves}
  is a software developer in the Department of Psychology at Stanford University, working in the Poldrack Lab. He is interested in building and improving tools for neuroimaging analysis and develops several open-source projects including fMRIprep and Nipype.
\end{IEEEbiography}

\begin{IEEEbiography}[{\includegraphics[width=1in,height=1.25in,clip,keepaspectratio]{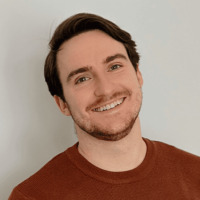}}]{Gregory Kiar} is a research scientist at the Child Mind Institute. Throughout his
  degrees in Biomedical Engineering, Greg has developed techniques to study
  biosignal data, a turn-key tool that allows researchers to generate maps of
  brain connectivity from diffusion MRI data, and techniques to assess and
  improve the stability of neuroscience research. Greg’s research bridges
  the fields of numerical analysis and brain imaging to evaluate and improve
  the trustworthiness of techniques used to study the brain.
\end{IEEEbiography}

\begin{IEEEbiography}[{\includegraphics[width=1in,height=1.25in,clip,keepaspectratio]{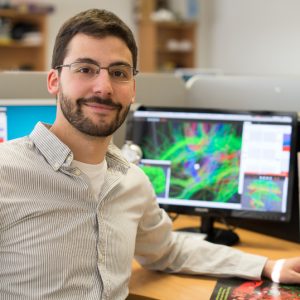}}]{Oscar Esteban}
  is a Research and Teaching Ambizione FNS Fellow at the Service of Radiology of the Lausanne University Hospital (CHUV) and the University of Lausanne. Oscar’s research aims at pushing the boundaries of neuroimaging — magnetic resonance imaging (MRI) mostly, — and by that, help other researchers advance our understanding of the human brain. In more specific terms, Oscar is currently developing tools that cater to researchers with “analysis-grade” data (see www.nipreps.org for more on this concept,) so they can focus on statistical modeling and inference. Perhaps, the flagship of these tools is fMRIPrep. This drive for the preprocessing step of the neuroimaging research workflow is justified by the concerning methodological variability that negatively contributes to reproducibility in the field. In particular, Oscar wants to improve the computational reproducibility of our results and minimize this methodological variability in the preprocessing step by standardizing workflows and reaching consensus implementations. In the longer term, Oscar’s vision is to contribute to uncovering the interplay of structure, function, and dynamics of brain connectivity using MRI.
\end{IEEEbiography}

\begin{IEEEbiography}[{\includegraphics[width=1in,height=1.25in,clip,keepaspectratio]{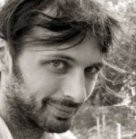}}]{Pierre Bellec}
  is the principal investigator of the laboratory for brain simulation and exploration (SIMEXP), as well as the scientific director of the Courtois Project on Neuronal Modelling (CNeuroMod), which uses human neuroimaging data to help train large artificial neural networks on a variety of cognitive tasks. He is also the director of Unité de neuroimagerie fonctionnelle at the “Centre de recherche de l’institut de gériatrie de Montréal” and an associate professor at the Psychology department at Université de Montréal. Dr Bellec is a senior fellow (”chercheur boursier senior”) of the ”Fonds de Recherche du Québec - Santé”.
\end{IEEEbiography}

\begin{IEEEbiography}[{\includegraphics[width=1in,height=1.25in,clip,keepaspectratio]{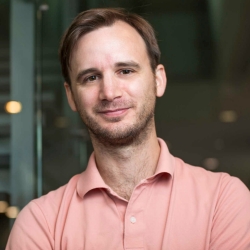}}]{Tristan Glatard} is Associate Professor in the Department of Computer Science and
  Software Engineering at Concordia University in Montreal, Canada
  Research Chair (Tier II) on Big Data Infrastructures for Neuroinformatics.
  Before that, he was a research scientist at the French National Centre for
  Scientific Research and Visiting Scholar at McGill University.
\end{IEEEbiography}

\vfill

\appendix

\section*{Numerical variability results for different smoothing kernel sizes}
\label{appendix:numerical_uncertainty}

\begin{figure*}
  \centering
  \uncertaintyMapDiscreteAll{0}
  \vspace*{-20pt}\hspace{15pt}
  \includegraphics*[width=.7\linewidth]{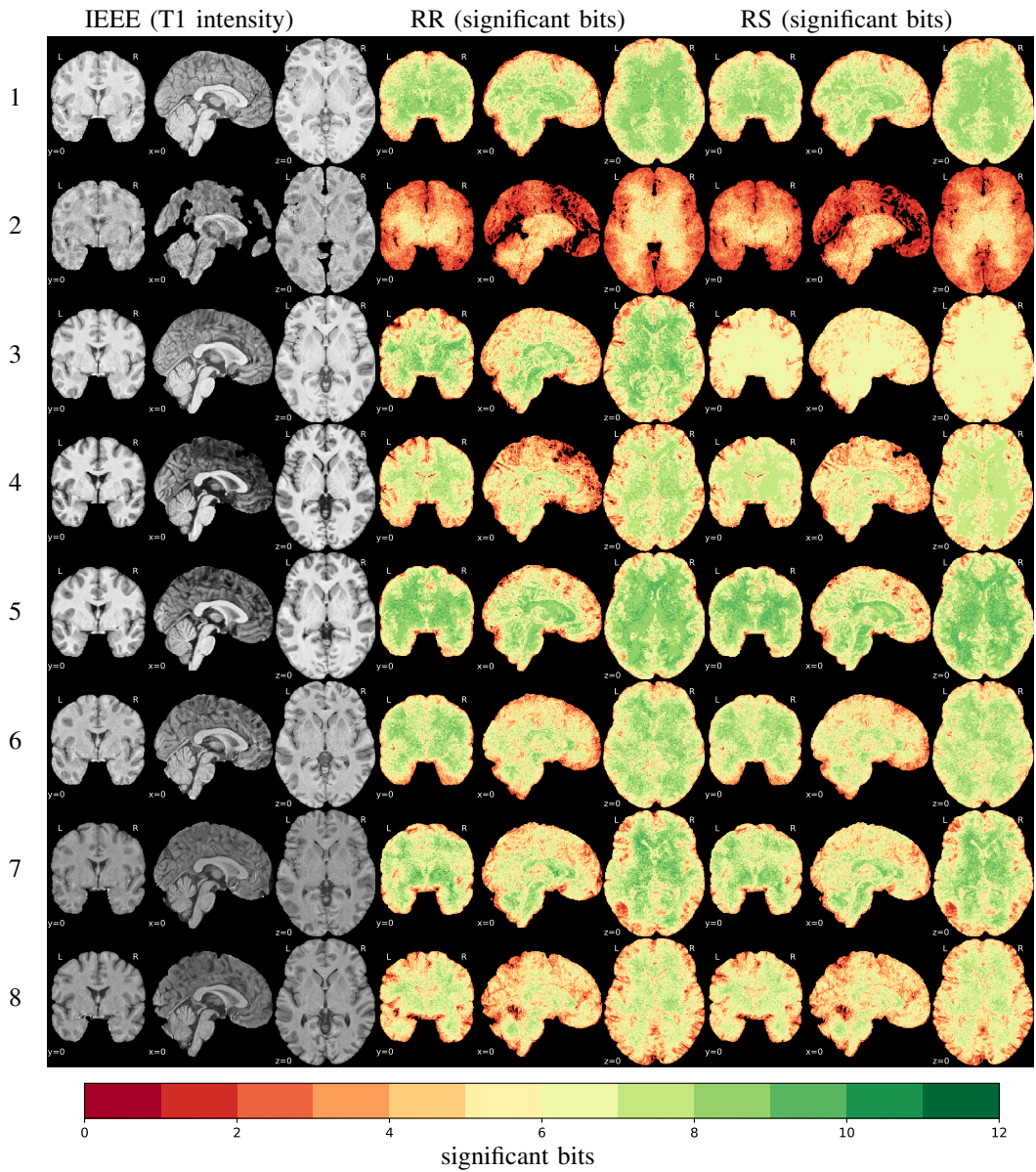} \\
  \vspace*{-5pt}
  significant bits
  \caption{Numerical variability measured for subjects 1 to 8 (from top to bottom) across n=30 perturbed samples, with FWHM\,=\,0\,mm. }
  \label{fig:uncertainty-maps-0mm-disc}
\end{figure*}

\begin{figure*}
  \centering
  \uncertaintyMapDiscreteAll{5}
  \vspace*{-20pt}\hspace{15pt}
  \includegraphics*[width=.7\linewidth]{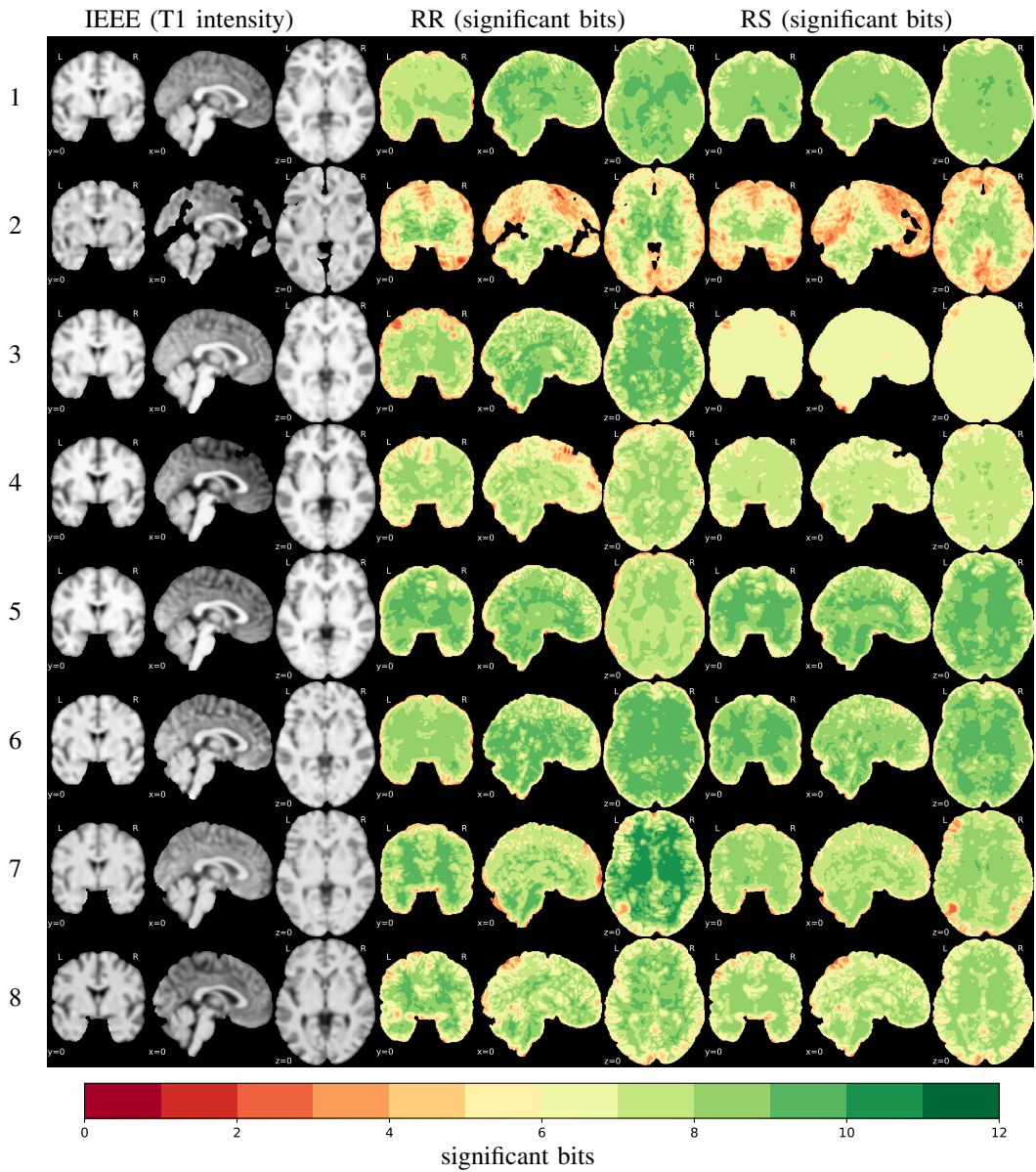} \\
  \vspace*{-5pt}
  significant bits
  \caption{Numerical variability measured for subjects 1 to 8 (from top to bottom) across n=30 perturbed samples, with FWHM\,=\,5\,mm. }
  \label{fig:uncertainty-maps-5mm-disc}
\end{figure*}

\begin{figure*}
  \vspace*{-2cm}
  \centering
  \uncertaintyMapDiscreteAll{10}
  \vspace*{-20pt}\hspace{15pt}
  \includegraphics*[width=.7\linewidth]{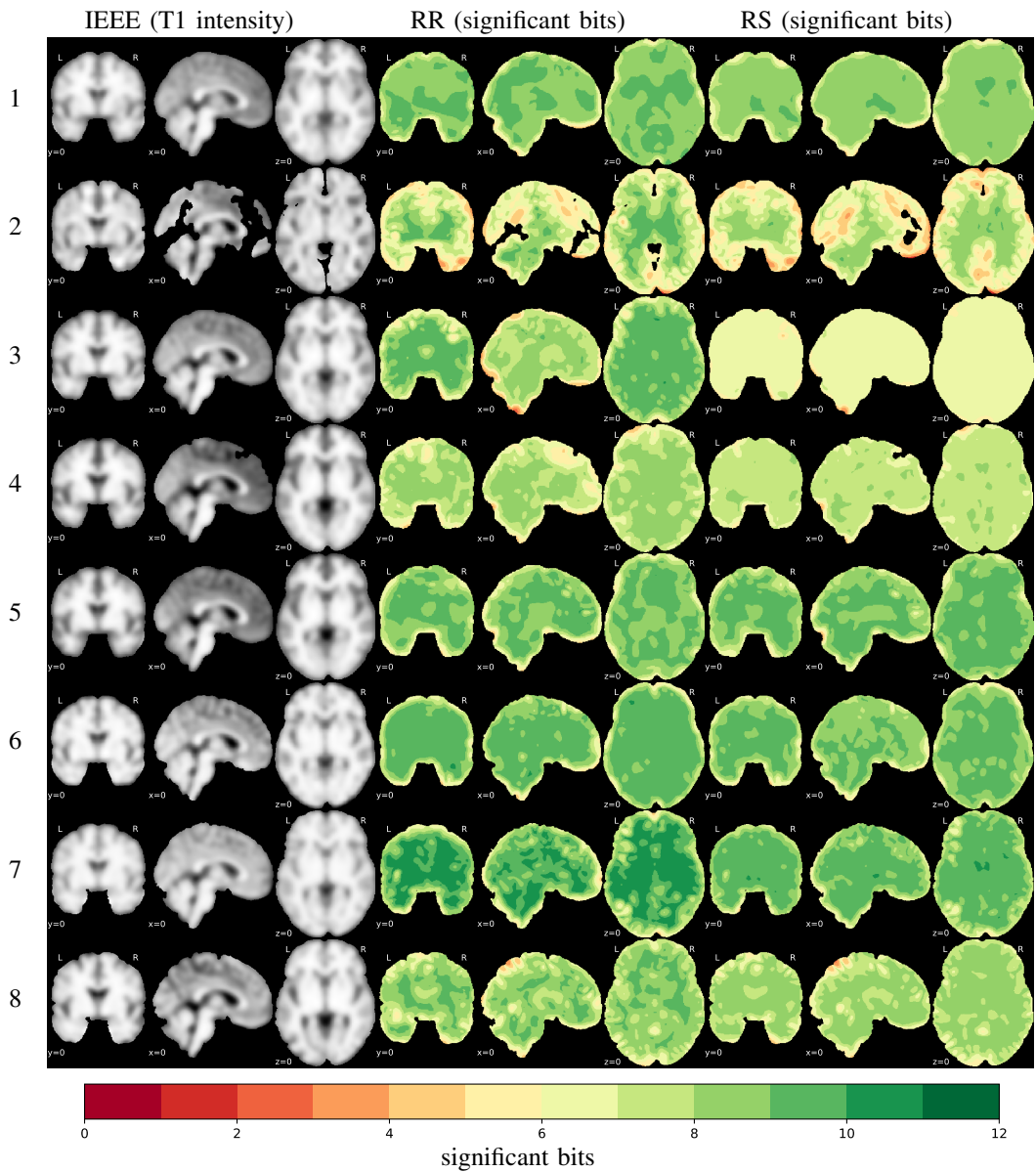} \\
  \vspace*{-5pt}
  significant bits
  \caption{Numerical variability measured for subjects 1 to 8 (from top to bottom) across n=30 perturbed samples, with FWHM\,=\,10\,mm. }
  \label{fig:uncertainty-maps-10mm-disc}
\end{figure*}

\begin{figure*}
  \vspace*{-2cm}
  \centering
  \uncertaintyMapDiscreteAll{20}
  \vspace*{-20pt}\hspace{15pt}
  \includegraphics*[width=.7\linewidth]{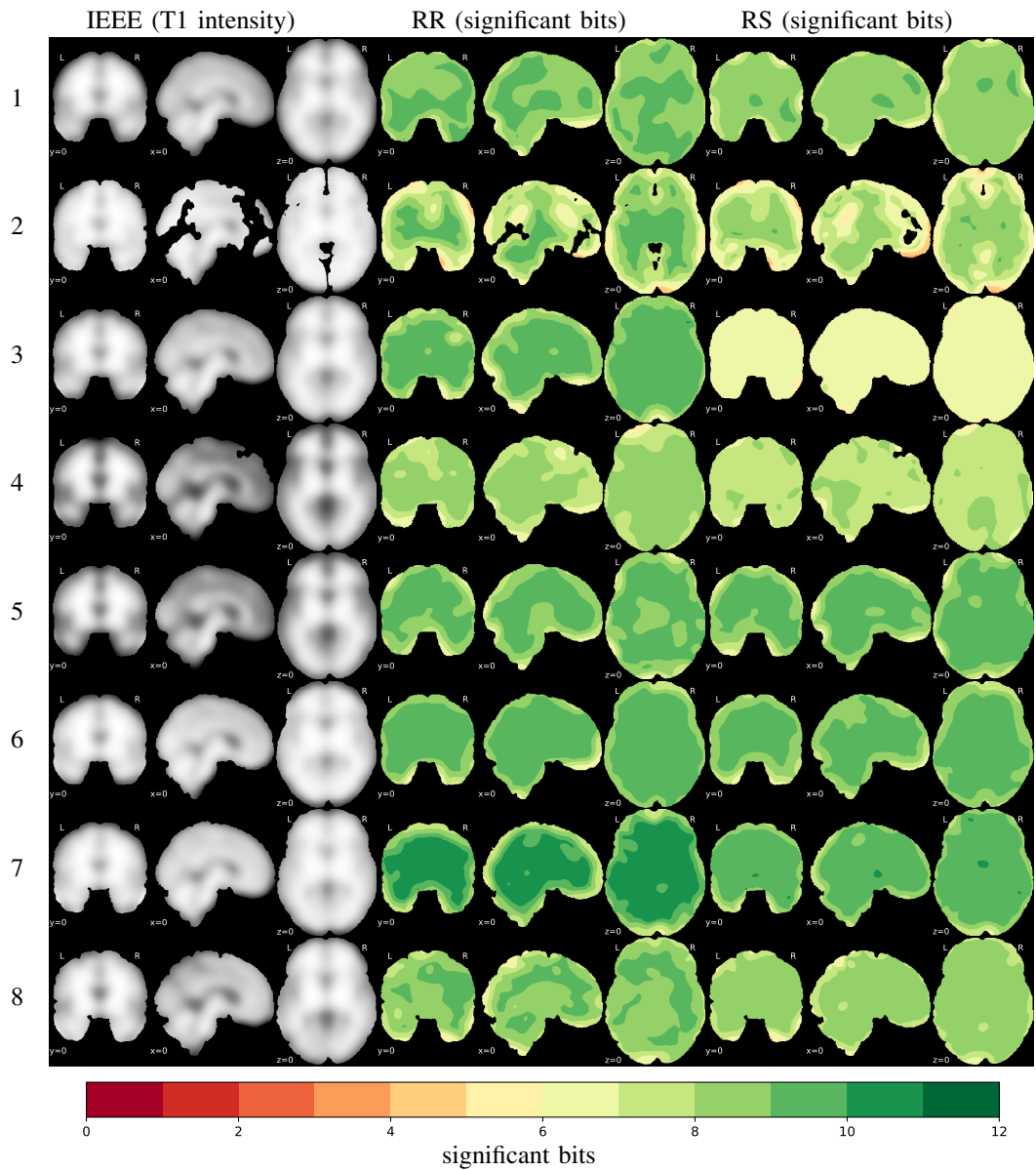} \\
  \vspace*{-5pt}
  significant bits
  \caption{Numerical variability measured for subjects 1 to 8 (from top to bottom) across n=30 perturbed samples, with FWHM\,=\,20\,mm. }
  \label{fig:uncertainty-maps-20mm-disc}
\end{figure*}

\end{document}